\documentclass[sigconf, review=false]{acmart}
\AtBeginDocument{%
  \providecommand\BibTeX{{%
    \normalfont B\kern-0.5em{\scshape i\kern-0.25em b}\kern-0.8em\TeX}}}

\usepackage{caption}
\usepackage{subcaption}
\usepackage{hhline}
\usepackage{pgfplotstable}
\usepackage{pgfplots}
\usepackage{marvosym}
\usepackage{xcolor}
\usepgfplotslibrary{groupplots} % Import the group plots library
\usepackage{blindtext}
\begin{document}

\setlength{\abovedisplayskip}{3.5mm}
\setlength{\belowdisplayskip}{3.5mm}

%% allowing the author to define a "short title" in page headers.
\title{SWAT: \underline{S}calable and Efficient \underline{W}indow \underline{A}ttention-based \underline{T}ransformers Acceleration on FPGAs}
%\title{FPGA based acceleration for Diagonal SDDMM}
%scalable windows for efficient sliding chunks
%% Of note is the shared affiliation of the first two authors and the
%% "author note" and "authornotemark" commands
\author{Zhenyu Bai, Pranav Dangi, Huize Li\textsuperscript{\Letter}, Tulika Mitra\textsuperscript{\Letter}}
\affiliation{%
   \institution{School of Computing, National University of Singapore, 119077, Singapore}
   \country{\{zhenyu.bai, dangi, huizeli\}@nus.edu.sg, tulika@comp.nus.edu.sg}
}
%\email{{zhenyu.bai, danji, huizeli}@nus.edu.sg, tulika@comp.nus.edu.sg }
\thanks{\textbf{Acknowledgement}: this research is partially supported by the National Research Foundation, Singapore under its Competitive Research Program Award NRF-CRP23-2019-0003 and AMD Research gift.}
\thanks{The corresponding authors are Huize Li and Tulika Mitra}

%%
%% By default, the full list of authors will be used on the page
%% headers. Often, this list is too long and will overlap

%%
%% The abstract is a summary of the work to be presented in the

\begin{abstract}
Efficiently supporting long context length is crucial for Transformer models. The quadratic complexity of the self-attention computation plagues traditional Transformers. Sliding window-based static sparse attention mitigates the problem by limiting the attention scope of the input tokens, reducing the theoretical complexity from quadratic to linear. Although the sparsity induced by window attention is highly structured, it does not align perfectly with the microarchitecture of the conventional accelerators, leading to suboptimal implementation. In response, we propose a dataflow-aware FPGA-based accelerator design, SWAT, that efficiently leverages the sparsity to achieve scalable performance for long input. The proposed microarchitecture is based on a design that maximizes data reuse by using a combination of row-wise dataflow, kernel fusion optimization, and an input-stationary design considering the distributed memory and computation resources of FPGA. Consequently, it achieves up to 22$\times$ and 5.7$\times$ improvement in latency and energy efficiency compared to the baseline FPGA-based accelerator and 15$\times$ energy efficiency compared to GPU-based solution.
\end{abstract}

\copyrightyear{2024} 
\acmYear{2024} 
\setcopyright{acmcopyright}\acmConference[DAC '24]{Proceedings of the 61th
ACM/IEEE Design Automation Conference (DAC)}{June 23--27, 2024}{San Francisco,
CA, USA}
\acmBooktitle{Proceedings of the 61th ACM/IEEE Design Automation Conference
(DAC) (DAC '24), June 23--27, 2024, San Francisco, CA, USA}
\acmPrice{15.00}
\acmDOI{10.1145/000000.0000000}
\acmISBN{xxxx}

\settopmatter{printacmref=false}
%\setcopyright{none}
%\renewcommand\footnotetextcopyrightpermission[1]{}
\maketitle
\pagestyle{plain}
%\pagenumbering{gobble}

\vspace{-2mm}
\section{Introduction}
\label{sec:intro}
Transformer-based models \cite{attention}, known for their self-attention mechanisms, are leading advancements in artificial intelligence. A typical transformer model contains linear layers, multi-head attention layers, and Feed-Forward Networks (FFN). 
The self-attention process involves first transforming each input token into the Query (Q), Key (K), and Value (V) vectors. Then it computes the dot products of Q and K for each token $S = Q \times K^T$, determining the similarity scores $S$ between them. These scores are then normalized using a softmax function to form probabilities $S' = SoftMax(S)$. Finally, the V vectors are multiplied by these probabilities and summed up to produce the final output $Z = S’ \times V$. This mechanism allows each token to contextually relate to \textbf{every} other token in the input, which is crucial for tasks requiring complex contextual understanding.

\begin{figure}
    \centering
    \resizebox{\linewidth}{!}{
    \begin{tikzpicture}
\begin{axis}[
    ybar stacked,   % Stacked horizontal bars
    bar width=8.3pt, % Width of the bars
    width= 5.5cm,
    height=3.2cm,
    xlabel={Input Length},
     xlabel style={
     xshift=0mm,
     yshift=-0.5mm,
    },
    title={FLOPs Breakdown},
    title style={
    yshift=-2mm, % Moves the title 2mm closer
    },
    legend style={at={(1.05,-0.7)},
    anchor=north,legend columns=-1},
    ylabel={Ratio}, % The label for the y axis
    ylabel near ticks,
    ylabel shift=-5pt,
    xtick=data,
    symbolic x coords={128,256,512,1024,2048,4096,8192,16384},% This option uses the x values from the data file
    x tick label style={rotate=45,anchor=east}, % This styles the tick labels
    ymin=0,  % Start y axis at 0
    ymax=1,  % End y axis at 1 (100%)
    enlarge x limits=0.15, % Add some breathing room around the data
    nodes near coords align={vertical}, % Align the nodes vertically
    ]

\pgfplotstableread[col sep=space]{data/FOP.data}\datatable
\pgfplotstablecreatecol[create col/expr={\thisrow{Liner}+\thisrow{Attention}+\thisrow{FFN}}]{sum}{\datatable}

% Plot the normalized stacks
\addplot+[ybar] table[x=input_length,y expr=\thisrow{Liner}/\thisrow{sum},col sep=space] {\datatable};
\addplot+[ybar] table[x=input_length,y expr=\thisrow{Attention}/\thisrow{sum},col sep=space] {\datatable};
\addplot+[ybar] table[x=input_length,y expr=\thisrow{FFN}/\thisrow{sum},col sep=space] {\datatable};

\legend{Liner, Attention, FFN}
\end{axis}

    \begin{scope}[xshift=4.4cm]
\begin{axis}[
    ybar stacked,   % Stacked horizontal bars
    bar width=8.3pt, % Width of the bars
    width= 5.5cm,
    height=3.2cm,
    title={MOPs Breakdown},
    title style={
    yshift=-2mm, % Moves the title 2mm closer
    },
    xlabel={Input Length},
    xlabel style={
     xshift=0mm,
     yshift=-0.5mm,
    },
    % legend style={at={(0.5,-0.35)},
    % anchor=north,legend columns=-1},
    %ylabel={Percentage}, % The label for the y axis
    xtick=data,
    symbolic x coords={128,256,512,1024,2048,4096,8192,16384},% This option uses the x values from the data file
    x tick label style={rotate=45,anchor=east}, % This styles the tick labels
    ymin=0,  % Start y axis at 0
    ymax=1,  % End y axis at 1 (100%)
    enlarge x limits=0.15, % Add some breathing room around the data
    nodes near coords align={vertical}, % Align the nodes vertically
    yticklabel=\empty,
    ]

\pgfplotstableread[col sep=space]{data/MOP.data}\datatable
\pgfplotstablecreatecol[create col/expr={\thisrow{Liner}+\thisrow{Attention}+\thisrow{FFN}}]{sum}{\datatable}

% Plot the normalized stacks
\addplot+[ybar] table[x=input_length,y expr=\thisrow{Liner}/\thisrow{sum},col sep=space] {\datatable};
\addplot+[ybar] table[x=input_length,y expr=\thisrow{Attention}/\thisrow{sum},col sep=space] {\datatable};
\addplot+[ybar] table[x=input_length,y expr=\thisrow{FFN}/\thisrow{sum},col sep=space] {\datatable};
\end{axis}
\end{scope}

    \end{tikzpicture}

    }
    \vspace{-5mm}
    \caption{Floating point operations (FLOPs) and memory operations (MOPS) breakdown for different input lengths}
    \label{fig:FOPs-MOPs}
    \vspace{-4.5mm}
\end{figure}
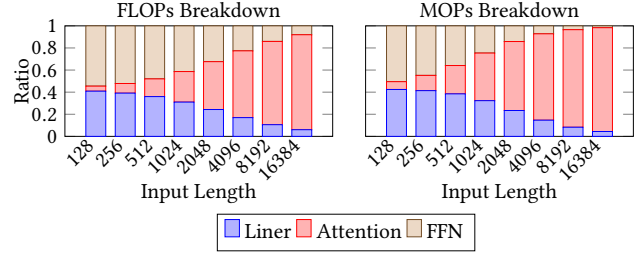

However, a notable impediment of this model is its quadratic complexity, which necessitates each sequence element to be compared against every other element. This complexity becomes particularly intractable in tasks with long contexts, such as document-level translation or long-form questions, due to the computational demands \cite{transformers-opt-survey}. This issue is illustrated in Figure~\ref{fig:FOPs-MOPs} where the floating point operations (FLOPs) and memory operations (MOPs) for attention computation grow with increasing input length and become a critical performance bottleneck.

\begin{figure}
    \centering
\begin{subfigure}[b]{0.22\textwidth}
    \centering
    \includegraphics[width=\textwidth]{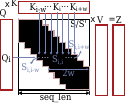}
    \vspace{-6mm}
    \caption{Window Attention}
    \label{fig:sliding-window}
\end{subfigure}
\hfill
\begin{subfigure}[b]{0.25\textwidth}
    \centering
    \includegraphics[width=\textwidth]{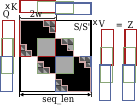}
    \vspace{-6mm}
    \caption{sliding chunks implementation}
    \label{fig:sliding-chunks}
\end{subfigure}

%     \begin{subfigure}[b]{0.2\textwidth}
%     \centering
%     \resizebox{\linewidth}{!}{
%         \begin{tikzpicture}
%         \pgfplotsset{scaled x ticks=false}
% \begin{axis}[
%   %title={Execution time (ms) per attention},
%   ylabel near ticks,
%   width= 5.5cm,
%     % title style={
%     % yshift=-2mm, % Moves the title 2mm closer
%     % },
%   height=4.0cm,
%   ylabel shift=-5pt,
%   xlabel shift=1pt,
%   xlabel={Input Length},
%   ylabel={Redundancy Ratio (\%)},
%   xmode=log, % If you want a logarithmic scale on the x-axis
%   xtick={512,1024,2048,4096,8192,16384},
%   xticklabels={512,1024,2048,4096,8192,16384},
%   x tick label style={rotate=45,anchor=east}, % This styles the tick labels
%   log ticks with fixed point,
%   grid=major,
% ]

% % Plot for 'dense'
% \addplot+ table[x=input_lenght, y=waste, col sep=space] {data/sliding-chunks-inefficieny.data};
% \end{axis}
% \end{tikzpicture}
% }
%     \caption{Ratio of redundant elements in sliding chunks}
%     \label{fig:sliding-chunks-inefficiency}
% \end{subfigure}
\vspace{-1mm}
\caption{Sliding window attention and its SOTA sliding chunks implementation}
\vspace{-5mm}
\end{figure}

To address this, the sliding window attention has been introduced \cite{longformer}. This method limits the attention of each token to a fixed subset of \textbf{adjacent} tokens, thereby reducing the theoretical complexity from quadratic to linear.
From the computational perspective, this method introduces a structured pattern of sparsity in the attention computation, as shown in Figure~\ref{fig:sliding-window} where each token attends to $w$ tokens before and after, forming a diagonal sparsity pattern of width $2w$. This sparsity manifests as a mask applied to the $S$ and $S'$ matrices, resulting in a Sampled Dense-Dense Matrix Multiplication between $Q$ and $K$, and a Sparse-Dense Matrix Multiplication between $S'$ and $V$. 
Despite the structured nature of this sparsity pattern, unlike dense operations, it is not perfectly aligned to be seamlessly filled into the vector/matrix math lanes of conventional accelerators. Instead, it requires fine-grain control of the microarchitecture for optimal performance that is not provided by existing accelerators. 
For instance, the Tensor Cores in Nvidia GPUs for accelerating dense matrix multiplications can only be accessed through higher-level programming models such as CUDA C++ API or C libraries, limiting micro-architecture level control. 
The current state-of-the-art implementation\footnote{Hugging Face's Longformer implementation, as per the original Longformer paper\cite{longformer}.}, known as \emph{sliding chunks}, addresses this limitation by dividing the sparse operation into smaller dense operations across chunks of width $2w$, as illustrated in Figure~\ref{fig:sliding-chunks}. 

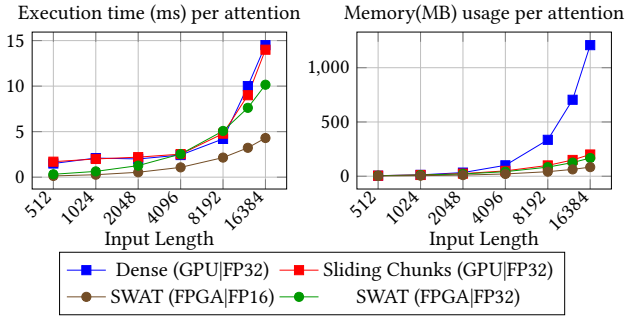
\begin{figure}
\resizebox{0.48\textwidth}{!}{
    \begin{tikzpicture}
\begin{axis}[
  title={Execution time (ms) per attention},
  ylabel near ticks,
  width= 5.5cm,
    title style={
    yshift=-2mm, % Moves the title 2mm closer
    },
  height=4.0cm,
  xmode=log,
  legend style={at={(1.1,-0.4)},anchor=north,legend columns=2},
  log ticks with fixed point,
  ylabel shift=-5pt,
  xlabel={Input Length},
    x tick label style={rotate=45,anchor=east}, % This styles the tick labels
    xtick={512,1024,2048,4096,8192, 16384},
  xticklabels={512,1024,2048,4096,8192, 16384},
  grid=major,
]

% Plot for 'dense'
\addplot[color=blue, mark=square*] table[x=input_lenght, mark=diamond, y=dense, col sep=space] {data/exec-time-motivation.data};
\addlegendentry{Dense (GPU|FP32)}
\addplot[color=red, mark=square*] table[x=input_lenght, y=sliding_chunks, col sep=space] {data/exec-time-motivation.data};
\addlegendentry{Sliding Chunks (GPU|FP32)}
\addplot[color=brown!60!black, mark=otimes*] table[x=input_lenght, y=ours_hf16, col sep=space] {data/exec-time-motivation.data};
\addlegendentry{SWAT (FPGA|FP16)}
\addplot[color=green!60!black, mark=otimes*] table[x=input_lenght, y=ours_fp32, col sep=space] {data/exec-time-motivation.data};
\addlegendentry{SWAT (FPGA|FP32)}

\end{axis}

\begin{scope}[xshift=5cm]
\begin{axis}[
  title={Memory(MB) usage per attention},
  ylabel near ticks,
  width= 5.5cm,
  height=4.0cm,
    title style={
    yshift=-2mm, % Moves the title 2mm closer
    },
  ylabel shift=-5pt,
  xlabel={Input Length},
  legend style={at={(0,0.8)},anchor=west,legend columns=-1},
  x tick label style={rotate=45,anchor=east}, % This styles the tick labels
  xmode=log,
  log ticks with fixed point,
  xtick={512,1024,2048,4096,8192, 16384},
  xticklabels={512,1024,2048,4096,8192, 16384},
  grid=major,
]

% Plot for 'dense'
%\addplot [color=orange, mark=triangle*] table[x=input_lenght, y=ours_streaming, col sep=space] {data/memory-motivation.data};
%\addlegendentry{SWAT(streaming)}

\addplot [color=blue, mark=square*] table[x=seq_len, y=dense, col sep=space] {data/memory-motivation.data};
%\addlegendentry{Dense}

% Plot for 'sliding chunks'
\addplot [color=red, mark=square*]table[x=seq_len, y=sliding_chunks, col sep=space] {data/memory-motivation.data};
%\addlegendentry{Sliding Chunks}

% Plot for 'ours'
\addplot [color=green!60!black, mark=otimes*] table[x=seq_len, y=ours, col sep=space] {data/memory-motivation.data};
%\addlegendentry{Ours}

\addplot[color=brown!60!black, mark=otimes*] table[x=seq_len, y=ours_fp16, col sep=space] {data/memory-motivation.data};

\end{axis}
\end{scope}

\end{tikzpicture}
}
    \caption{Execution time and memory usage of existing approaches Dense and Sliding Chunks compared to SWAT}
    \vspace{-6mm}
    \label{fig:motivation}
\end{figure}

Yet, this approach leads to redundant computations in the form of the overlapping regions (gray regions in figure~\ref{fig:sliding-chunks}) and corner areas (dashed regions in the figure) of each chunk. The ratio of these redundant computation is given by $\frac{1}{2} - \frac{1}{4|chunks|}$ where $|chunks|$ is the number of chunks. This ratio increases and approaches rapidly to $\frac{1}{2}$, i.e., 50\% redundancy.
Moreover, eliminating these redundant calculations is challenging as the correctness of the results must be ensured, which further increases the computational overhead. 
In Figure~\ref{fig:motivation}, We compare the execution time and memory usage\footnote{The experimental setup will be presented in Section 5.} of the sliding chunks with the naïve dense operations approach on an AMD MI210 GPU. The result indicates that while the sliding chunks approach significantly reduces memory usage, the computational time remains similar to the dense method, primarily due to the redundant computation but also due to the overhead for increased frequency of small kernel launches on GPU. 

\paragraph{\textbf{Motivation \& Contribution:}}
We aim to refine the implementation of sliding window attention by focusing on the efficient computation of its structured, yet imperfectly aligned sparsity, which requires precise control over computation and memory operations.  We propose a novel hardware design using Field-Programmable Gate Arrays (FPGAs). FPGAs are favored over ASICs because their programmability allows for the support of various attention mechanisms such as global attention and random attention, enhancing the model accuracy across different tasks. Additionally, FPGAs are readily available on cloud platforms, e.g., Microsoft Azure\cite{ms-asure-fpga} and Amazon AWS, offering a cost-effective deployment solution. %But our solution can also be easily instantiated as an ASIC accelerator. 

The implementation of the sliding window attention has to consider the computation workload and how it is mapped onto the distributed memory and computation resources of the FPGA fabric for optimal performance. Therefore, we deeply analyze the data flow of the sliding window attention, which reveals that a combination of window attention, row-major dataflow, and kernel fusion can significantly enhance off-chip transfer efficiency, ensuring each data item is loaded just once. We propose employing a fixed-size First-In-First-Out (FIFO) buffer to manage the sliding window input, leading to an input-stationary data flow. Here, the input data remains in the buffers after being loaded while necessary computational resources are placed around them by the micro-architecture design. This approach better utilizes the on-chip memory blocks' bandwidth and minimizes on-chip data movement, aligning well with the distributed memory and computation units of the underlying FPGA fabric and therefore achieving higher performance. 

As demonstrated in Figure~\ref{fig:motivation}, SWAT exhibits linear scaling of memory use with input length. 
SWAT achieves 6$\times$ energy efficiency to conventional GPU-based solutions for comparable execution time for input length below 8K tokens and shows superior performance for longer input.  When compared to a baseline FPGA accelerator, SWAT achieves 22$\times$ and 5.7$\times$ improvements in latency and energy efficiency, respectively (with 16384 tokens). Moreover, SWAT shows better scalability compared to both GPU and FPGA solutions.

\section{Background \& Related works}
%\vspace{-2mm}
\subsection{Efficient Transformers}
%\vspace{-1mm}
Efficient Transformers~\cite{efficient-survey} are the variants of the vanilla Transformer model, aiming at improving the computational efficiency. 
Two main strategies underlie these improvements.
The first seeks to approximate the traditional SoftMax attention mechanisms with algorithms of lower computational complexity. For instance, FNet~\cite{fnet} and Linformer~\cite{linformer} substitute the standard attention calculations with Fourier Transforms and linear projections, respectively. 
The second pathway, sparse attention, aims to reduce attention operations while preserving the SoftMax-based attention formulation.

\vspace{-1.5mm}
\subsection{Sparse attention}
\vspace{-0.5mm}
Sparse attention can be further classified into two categories. Dynamic mathods~\cite{sanger,dota,fact} evaluate or predict attention scores in real-time, prioritizing the most significant ones for subsequent computation. These methods, while adaptable, introduce irregular sparsity patterns that hinder efficient computation. In contrast, static methods~\cite{longformer,bigbird,butterfly-model} pre-define attention patterns, achieving structured sparsity at the cost of some accuracy. The structured sparsity can be leveraged for predictable performance gains with static optimization and dedicated hardware support. 

Sliding window attention~\cite{longformer}, is the key component of nearly all static sparse attention approaches. This technique limits each token's attention to a predetermined number of adjacent tokens, based on the research findings that show the substantial impact of the local context within the attention mechanism~\cite{local-attention-vision, VITCOD}. 

\vspace{-1.5mm}
\subsection{Accelerators for static structured attention}
\vspace{-0.5mm}
ASIC-based accelerator SALO~\cite{salo}, designed explicitly for the Longformer model, utilizes structured sparsity in window attention with a 2D square systolic array. 
However, it is limited to the basic Longformer setup. %Research suggests that integrating random attention and multiple global tokens is essential for consistent model performance on different tasks\cite{ViL,bigbird}, which is intractable to be extended from the SALO's ASIC design.
Furthermore, the systolic array's square structure requires square tiling of the input and the intermediate matrices. This tiling is suboptimal for row-wise SoftMax operations, necessitating supplementary computations outside the accelerator's capabilities.

Butterfly~\cite{butterfly} is an FPGA-based accelerator for efficient transformer models that utilizes a static butterfly sparsity pattern. Notably, this pattern can be approximated through Discrete Fourier Transformations. However, this FFT-based approximation of the original SoftMax attention has not consistently demonstrated reliable accuracy across various downstream tasks. Butterfly attempts to mitigate this by combining vanilla SoftMax attention layers with FFT-based attention layers. Our analysis in Section~\ref{sec:accuracy-butterfly} reveals that incorporating at least one conventional attention mechanism is crucial for acceptable accuracy. Unfortunately, this traditional attention's quadratic complexity leads to suboptimal performance for the accelerator when managing long input sequences.

\begin{figure}[h]
\vspace{-2mm}
\centering
\begin{subfigure}[b]{0.48\linewidth}
    \centering
    \includegraphics[width=\textwidth]{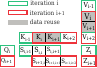}
    \vspace{-4mm}
    \caption{Data-reuse}
    \label{fig:data-reuse}
\end{subfigure}
\hfill
\hfill
\begin{subfigure}[b]{0.44\linewidth}
    \centering
    \includegraphics[width=\textwidth]{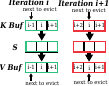}
    \vspace{-4mm}
    \caption{Fixed-length FIFO}
    \label{fig:fifo-buf}
\end{subfigure}
\vspace{-1mm}
\caption{Data management for window attention}
\vspace{-5mm}
\end{figure}

\section{Dataflow Analysis \& Optimization}
\label{sec:dataflow}
%\vspace{-1.5mm}
\subsection{Enabling kernel fusion}
%\vspace{-1mm}
\label{sec:kernel-fusion}
The standard implementation of Transformer models involves a sequential three-step computation- $QK$ multiplication, SoftMax, and $SV$ multiplication- primarily due to the row-wise data dependency of the SoftMax operation. 
As on-chip memory is typically insufficient for handling the entire computation in one go, each step is broken down into smaller tile-wise operations, leading to redundant off-chip data transfers for loading and storing the tile-wise intermediate results (the tiles of $S$ and $S'$).

Kernel fusion~\cite{flashattention}, as an optimization technique, aims to consolidate these steps into a single operation for each input tile, thereby reducing off-chip data transfers. However, the inherent row-wise dependency in the SoftMax operation presents a significant challenge to this fusion. By reinterpreting the SoftMax operation, we can divide it into two components: the numerator that does not depend on the other elements of the row and the denominator that depends on the sum of the exponential of all elements of the same row. By viewing the denominator as a scaling factor, it can be placed after the third step $Z = S' \times V$  as shown in Equation~\ref{eq:fusion}. This restructuring allows for the fusion of the three operations into a unified \textbf{row-wise} kernel.

\vspace{-7mm}
\begin{equation}
    \begin{split}
        Z_{i,j} & =  \sum_{n = 0}^{H} S'_{i,n} V_{n,j} = \sum_{n = 0}^{H} \frac{exp(S_{i,n})}{\sum_{l = 0}^{H}exp(S_{i,l})}  V_{n,j}\\
          & = ( \frac{1}{\sum_{l = 0}^{H}exp(S_{i,l})} ) ( \sum_{n = 0}^{H} exp(S_{i,n})  V_{n,j}) \\
          %& = \alpha ( \sum_{n = 0}^{H} exp(\sum_{m=0}^{H} Q_{i,m} \cdot K^T_{m,n} )  V_{n,j})
    \end{split}
    \label{eq:fusion}
\end{equation}
\vspace{-6mm}
\subsection{Row-major dataflow \& Data reuse}
\vspace{-0.5mm}
In the standard three-step computation of transformers, independent execution of each step limits the benefits gained from tiling strategies or dataflow optimization. However, in the context of sliding window attention, these three computations share a common sparsity pattern, as depicted in Figure~\ref{fig:sliding-window}. When considering the attention computation for a given input row of $Q$, say $Q_i$, and the subsequent row vector $Q_{i+1}$, we observe significant data reuse of the attended rows of $K$ (columns of $K^T$) and $V$, as shown in Figure~\ref{fig:data-reuse} for the window width $w=1$. The most effective way to harness this data reuse is by adopting a row-major dataflow. Although kernel fusion \emph{postpones} the row-wise dependencies of SoftMax, it does not \emph{eliminate} them. A row-major approach, therefore, becomes advantageous, minimizing the memory needed for storing intermediate results $S$ and $S'$, which now can be stored in on-chip memories.

To capitalize on this data reuse opportunity, SWAT employs fixed-size on-chip FIFO buffers for the $K$ and $V$ inputs, while the $Q$ input changes for each row. This setup, illustrated in Figure~\ref{fig:fifo-buf}, features a buffer with a moving pointer that indicates the next element to be replaced, ensuring data is loaded exactly once and achieving 100\% off-chip memory transfer efficiency.

\vspace{-1.5mm}
\subsection{Input-Stationary dataflow for FPGA}
\vspace{-0.5mm}
The previous sections discussed SWAT's dataflow at the algorithmic level. Now, it's important to consider how this dataflow is mapped onto the FPGA's microarchitectural design. The following key aspects are considered.
First, FPGAs have distributed memory (BRAMs and URAMs) and computing elements (LUTs, DSP slices) across the chip. Secondly, by utilizing fixed-size FIFO buffers for input data, as exhibited in Figure~\ref{fig:fifo-buf}, input data mostly remains stationary. Finally, kernel fusion ensures a consistent pairing of $(K_j, V_j)$ for each Key/Value row $j \in [i-w,i+w]$. This coherency is apparent both in the fusion equation (see Equation~\ref{eq:fusion}) and the FIFO eviction process (Figure~\ref{fig:fifo-buf}). From the algorithm level, this is because the same input sequence indexes $Key$ and $Value$ matrices according to the self-attention mechanism.

\begin{figure}
    \centering
    \vspace{-2mm}
    \includegraphics[width=0.85\linewidth]{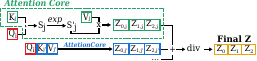}
    \vspace{-1mm}
    \caption{input-stationary dataflow}
    \vspace{-6mm}
    \label{fig:row-dataflow}
\end{figure}

Consequently, we adopt an \emph{input-stationary} dataflow. In this design, input data remain in their respective buffers, and computational units are positioned nearby. This is different from conventional accelerator designs, where the data is brought to the computational units. Figure~\ref{fig:row-dataflow} illustrates this dataflow for one row of input $Q$ within the input stationary paradigm. An \emph{Attention Core}—our terminology for the minimal computational unit—consists of a buffer holding one row of $K$ ($K_j$), and one row of $V$ ($V_j$). Upon the arrival of a new row of $Q$, denoted as $Q_i$, the multiplication with $K_j$ is performed locally within each Attention Core: $S_{i,j} = Q_i \cdot K_j$. Subsequently, the numerator of the SoftMax computation is performed: $S'_{i,j} = exp(S_{i,j})$ according to the kernel fusion. For the multiplication of $S'$ with $V$, we adhere to the input-stationary paradigm, where each $S'$ element multiplies with the corresponding $V$ row stored in the \textbf{same} attention core. This operation yields one slice of $Z$ per attention core. The slices produced by all attention cores are summed up outside of the attention cores to form the final result  $Z$.

\vspace{-1.5mm}
\subsection{Dataflow compatibility for ASIC}
\vspace{-0.5mm}
The dataflow optimization techniques we have developed, particularly row-major dataflow and kernel fusion, are also applicable to ASIC-based implementations, which, unlike FPGAs, are not constrained by the distribution of computation and memory resources. While ASICs can potentially offer superior performance, they lack the flexibility of FPGAs, which is crucial for adapting to the evolving landscape of Transformer models.

%Implementing our proposed dataflow on GPUs presents significant challenges due to both programming models and architectural limitations. Nvidia GPUs, for instance, rely on programming models like CUDA C++ API and various CUDA libraries (cuBlas, cuSparse, cuDNN, etc.). While the latter high-level libraries are optimized for standard dense/sparse operations, they are not suited for customization. This is the reason behind sliding chunks's design choice. On the other hand, CUDA C++ provides more detailed control over execution but adopts a thread-centric view. This perspective means that programmers are discouraged from micromanaging how threads are scheduled and allocated to the computation and memory units. Instead, workload assignment and distribution are dynamically managed by the hardware. This inherent characteristic of GPU architecture makes it challenging to implement our dataflow optimization, which requires a more granular and controlled approach to data and computation management.

\begin{figure*}[h]
    \centering
    \includegraphics[width=0.93\textwidth]{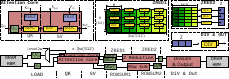}
    \caption{SWAT Microarchitecture design}
    \label{fig:archi-design}
    \vspace{-2.7mm}
\end{figure*}

\section{Architecture design}
%\vspace{-1mm}
\label{sec:archi}

Figure~\ref{fig:archi-design} shows the architecture design of SWAT by following the dataflow design outlined in the previous section. 
The architecture makes use of a pipeline execution to improve resource usage efficiency. The functions of the pipeline stages are as follows:

\emph{LOAD Stage}: 
Data from the main memory is fetched and loaded into the $K/V$ buffers of the attention cores. For the standard window width configuration ($2w=512$), 512 attention cores are instantiated.
Each $K/V$ buffer uses one BRAM block, storing a full row of $K$ or $V$ of size $H$(head dimensionality). 
According to the $K/V$ buffer replacement policy, the entire $K/V$ buffer of \textbf{one} attention core is refreshed per attention of one row. The selection signal is computed by the row index $i$ modulo the window size according to the FIFO policy. The $Q$ row is loaded during this stage and distributed across all attention cores. 

\emph{QK Stage}: This stage calculates the dot product between the $K$ row and the $Q$ row in the attention cores. Due to FPGA constraints, the FP16 multiply-accumulate (MAC) operation is pipelined at an Initial Interval (II) of 3 cycles. Forcing the MAC to be pipelined at fewer cycles will significantly increase resource usage. 

\emph{SV Stage}:
Following the QK stage, the SV stage computes the exponential of the $S$ values and multiplies these with the corresponding $V$ elements within the \textbf{same} attention core, generating a slice of $Z$ per attention core, stored in $ZBuf$. The FP16 multiplications are executed over an II=3 pipeline. While a lower II is feasible, it does not improve overall performance due to the II=3 of QK stage and would lead to increased resource usage for pipelining.

\emph{Z Reduction}: 
This two-phase stage sums the individual $Z$ slices from each attention core to form the complete output $Z$ vector. For a standard configuration of $H=64$, parallel accumulation over $H$ channels (because $Z$ has $H$ elements) would result in a stage duration of approximately $3 \times 2w$ which is 8x that of QK and SV stages of $3 \times H$ cycles. To maintain pipeline balance, the reduction is split into two substages, ZRED1 and ZRED2. In ZRED1, $Z$ slices are grouped by each $H$ of them and processed with $H$ accumulation channels per group, which results in approximately an overall latency of $3\times H$ cycles. ZRED2 then combines the outputs from ZRED1 into the final $Z$ vector.

\emph{Row sum}: Operating in parallel to Z reduction, the Row Sum stage computes the sum of $S'$ values from the attention cores. With $2w$ elements of $S'$, this stage employs a similar two-stage approach as Z reduction for timing balance, comprising ROWSUM1 and ROWSUM2.

\emph{Division and Output}: 
The final stage divides each $Z$ element by the corresponding sum of the $S'$ row, as per the post-fusion algorithm. The division is pipelined at a 2-cycle interval because better throughput is unnecessary. The output vector is then written back to HBM or DRAM. 

Table~\ref{tab:pipeline-stages-timing} presents the timing for each pipeline stage based on the Xilinx Vitis HLS synthesis tool report. The design uses half-precision 16-bit floating-point data, with default settings of head dimension $H=64$ and window width $2w=512$. The overall pipeline is well balanced and timed at 201 cycles, predominantly due to the longer stage, $QK$.

\begin{table}[h]
    \centering
    \vspace{-2mm}
    \caption{The timing (in cycles) of the pipeline stages}
    \vspace{-1mm}
    \resizebox{0.8\linewidth}{!}{
    \begin{tabular}{|c|c|c|c|c|c|}
    \hline
         ~      & ~     & ~     & ZRED1 & ZRED2 &\\
    \hhline{|~~~|~|~|~|}
         LOAD   & QK    & SV    & 195   & 66    &  DIV\&OUT  \\
    \hhline{|~~~|-|-|~|}
        66     & 201   & 197    & ROWSUM1 &ROWSUM2 & 179 \\
    \hhline{|~~~|~|~|~|}
         &       &       & 195   & 27        &   \\
    \hline
    \end{tabular}
    }
    \label{tab:pipeline-stages-timing}
    \vspace{-2mm}
\end{table}

\vspace{-4mm}
\subsection{Parameterized design}
SWAT's architecture integrates basic window attention with additional attention mechanisms to improve accuracy across various tasks. One such mechanism is \emph{global attention}, which designates important global tokens to be attended by all input tokens. Models like Longformer~\cite{longformer} and ViL~\cite{ViL} have demonstrated the effectiveness of global attention in enhancing accuracy for text classification and vision tasks, respectively. Another mechanism, \emph{random attention}, introduced by the BigBird model~\cite{bigbird}, generally improves model accuracy by incorporating randomly (but statically) selected additional tokens for each input token to attend to.

\begin{figure}[!h]
    \centering
    \vspace{-3mm}
    \includegraphics[width=0.9\linewidth]{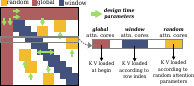}
    \caption{Parameterized Design of SWAT}
    \vspace{-4mm}
\label{fig:attention-addons}
\end{figure}

As illustrated in Figure~\ref{fig:attention-addons}, SWAT supports these mechanisms through design-time parameters: the indices of the random and global attention tokens, as well as the width of the sliding window, are set as synthesis parameters. SWAT's architectural design can adapt to these additional attention patterns with minimal changes. Specific subsets of attention cores are allocated for computing global and random attention with respect to the parameters. Attention cores dedicated to global attention have fixed K and V buffers, aligning with the consistent nature of global tokens. These buffers are pre-loaded prior to the attention computation, minimizing performance impact. In contrast, attention cores handling random attention update their K and V buffers dynamically, which increases the latency of the LOAD stage to 195 cycles from the initial 66. However, thanks to the pipelined design of our system, this increase in latency does not hamper overall execution speed.

\vspace{-1.5mm}
\subsection{FPGA resources utilization}
\vspace{-0.5mm}
Table~\ref{tab:resource-usage} provides a detailed account of resource usage on the Alveo U55C FPGA post-synthesis. We present four configurations: the standard Longformer setup of pure window attention with FP16 datatype and 512 attention cores; the BigBird configuration of 192 sliding window tokens, 192 random attention tokens, 128 global tokens, i.e. total 512 tokens per row with FP16; the same BigBird configuration but with dual pipelines for parallel processing two heads (which also demonstrates the potential of handling 1024 tokens per row in different attention configurations); and an FP32 version for later comparative analysis with GPUs.

\begin{table}[h]
    \centering
    \vspace{-3mm}
    \caption{Resources usage on U55C/VCU128}
    \vspace{-1mm}
    \resizebox{0.87\linewidth}{!}{
    \begin{tabular}{|c|c|c|c|c|}
    \hline
        Design &DSP & LUT   & FF    & BRAM   \\
    \hline
        FP16 (512 attn) & 19\% & 38 \%    & 11\%    & 25\%     \\
        FP16 (BigBird 512 attn)       & 19\% & 33 \%    & 11 \%      & 25\%     \\
        FP16 (BigBird 2 x 512 attn)       & 38\% & 66 \%    & 22 \%      & 50\%     \\
    \hline
        FP32 (512 attn) & 49\% & 67\% & 23\% &25\% \\
        
    \hhline{|=|=|=|=|=|}
        Butterfly (FP16, 120-BE)      & 32\%     & 79\%    & 63\%   & 49\% \\
    \hline
    \end{tabular}
    }
    \vspace{-5mm}
    \label{tab:resource-usage}
\end{table}
\section{Evaluation}
%\vspace{-1mm}
\subsection{Butterfly accelerator baseline}
\vspace{-0.5mm}
The Butterfly Accelerator\cite{butterfly} is the only FPGA-based accelerator for static sparse attention--the butterfly sparsity\cite{butterfly-model}-- and serves as our baseline.
It incorporates two key hardware components: the FFT-BTF (Fast Fourier Transform-Butterfly) engine for \textbf{approximating} the standard SoftMax attention using Fourier transform; and the ATTN-BTF (Attention-Butterfly) engine that behaves just as the standard SoftMax attention. The FFT-BTF offers increased speed at the expense of some accuracy, whereas the ATTN-BTF ensures accuracy with slower operation. The hybrid use of FFT and SoftMax layers in Butterfly's software model is tuned for specific datasets to achieve a balance between speed and accuracy through design space exploration. However, the performance study of the Butterfly Accelerator in~\cite{butterfly} focuses only on the full-FFT version.

\vspace{-1.5mm}
\subsection{Accuracy comparison with Butterfly}
\vspace{-0.5mm}
\label{sec:accuracy-butterfly}
To draw a fair comparison with SWAT, we delve into the accuracy-performance tradeoff in Butterfly's adaptable design. We evaluated model accuracies using the Long-Range Arena (LRA) benchmark datasets~\cite{LRA}, which are tailored for efficient transformer models.
Table~\ref{tab:accuracy-LRA} shows the \textbf{accuracy advantage} of SWAT implementations over the purely FFT-layered Butterfly model, particularly in vision tasks. Additionally, we observe an accuracy improvement in the Butterfly model when replacing one or two last FFT layers with traditional SoftMax attention layers (respectively noted as configuration BTF-1 and BTF-2), underscoring the accuracy benefit of incorporating even a single layer of traditional SoftMax attention. However, Longformer and Bigbird still show better average accuracy compared to BTF-1 and BTF2, especially in vision tasks. For the accuracy consideration, we will use BTF-1 and BTF-2 in the performance analysis in the next section.

\begin{table}
    \centering
    \caption{Accuracy gain of window attention-based models (Longformer and BigBird supported by SWAT) and baseline Butterfly models with one or two layers (BTF-1, BTF-2) replaced by the vanilla SoftMax attention on LRA datasets compared to Butterfly's full-FFT attention}
    \vspace{-1mm}
    \resizebox{.45\textwidth}{!}{
    \begin{tabular}{|c|c|c|c|c|c|}
        \hline
            & \multicolumn{2}{c|}{Vision based} & \multicolumn{2}{c|}{Text based} & \\
        \hhline{|~|--|--|}
        Model & Image & PathFinder & Text & ListOps &AVG. \\
        \hhline{|=|=|=|=|=|=|}
        Longformer	         &+15.26\%	&+3.03\%	&+0.17\%	&+1.61\%	&+5.02\% \\
        Bigbird	            &+13.87\%	&+8.16\%	&+1.34\%	&+2.03\%	&+6.35\% \\
        \hline
        BTF-1	 &+6.26\%	&+2.85\%	&+0.01\%	&+2.4\%	     &+3.01\% \\
        BTF-2 &+8.95\%	&+2.14\%	&+1.05\%	&+2.42\%	&+3.64\% \\
        \hline
    \end{tabular}
    }
    \vspace{-3mm}
    \label{tab:accuracy-LRA}
\end{table}

\begin{figure}
    \centering
    \begin{minipage}{0.44\linewidth}
        \captionsetup{type=table}
        \caption{Top-1 accuracy of PixelFly (butterfly model) against ViL (supported by SWAT) on ImageNet-1K~\cite{imagenet}}
        \centering
        \resizebox{1.03\linewidth}{!}{
        \begin{tabular}{|c|c|c|}
        \hline
        Model                    & Params & Top-1  \\
        \hline  
        ViL-Tiny	             &6.7M	     &76.7\% \\
        %Pixelfly-Mixer-S	 &5.9	     &72.6 \\
        Pixelfly-M-S	 &5.9M	     &72.6\% \\
        \hline
        ViL-Small	             &24.6M	     &82.4\% \\
        Pixelfly-V-S	     &16.9M	      &77.5\% \\
        Pixelfly-M-B     & 17.4M         & 76.3\%\\
        Pixelfly-V-B	         &28.2M	      &78.6\% \\
        %Pixelfly-B	         &28.2M	      &78.6 \\
        \hline
        ViL-Med	                &39.7M	      &83.5\% \\
        \hline
        \end{tabular}
        }
        \label{tab:vil-vs-pixelfly}
    \end{minipage}
    \begin{minipage}{0.55\linewidth}
        \centering
        \resizebox{\linewidth}{!}{
        \begin{tikzpicture}
        \pgfplotsset{every tick label/.append style={font=\footnotesize}}
\begin{axis}[
  ymin=0, ymax=45,
  ylabel near ticks,
  ylabel={Normalized Speedup},
  ybar,
  width= 4.6cm,
  height=4cm,
  bar width=5pt,
  ylabel shift=-6pt,
  %nodes near coords,
  xlabel={Input Length},
  xtick=data,
  x tick label style={rotate=0,anchor=north,yshift=1.6mm},
  symbolic x coords={128,256,512,1024,2048,4096,8192,16384},% This option uses the x values from the data file
  grid=major,
  xticklabel style={
    /pgf/number format/fixed,
        /pgf/number format/precision=5
  },
  title style={
    yshift=-3mm, % Moves the title 2mm closer
  },
  xlabel style={yshift=2.5mm},
  label style={font=\footnotesize},
  xmajorgrids=false,
  xminorgrids=false,
  ytick={5,10,15,20,25,30,35,40,45},
  yticklabels={5x,10x,15x,20x,25x,30x,35x,40x,45x},
  scaled x ticks=false,
  legend style={at={(0,0.86)},anchor=west,legend columns=1,font=\footnotesize,
     inner xsep=2pt, % Adjust horizontal padding inside the legend
            inner ysep=0pt, % Adjust vertical padding inside the legend
            outer xsep=0pt, % Adjust horizontal spacing outside the legend
            outer ysep=0pt, % Adjust vertical spacing outside the legend
            cells={anchor=west}, % Align legend text to the left
    },
    legend image code/.code={
                \draw [/tikz/.cd, bar width=3pt,yshift=-0.25em,bar shift=0pt]
                plot coordinates {(0cm,0.5em)};
            },
    xtick style={draw=none},
]

\addplot+ table[x=seq_len, y=speedup1softmax, col sep=space] {data/speedup.data};
\addlegendentry{SWAT vs. BTF-1}

\addplot+ table[x=seq_len, y=speedup2softmax, col sep=space] {data/speedup.data};
\addlegendentry{SWAT vs. BTF-2}

\end{axis}
\end{tikzpicture}
}   
\vspace{-7mm}
        \caption{Speedup (times) of SWAT over Butterfly versions}
        \label{fig:performance}
    \end{minipage}
    \vspace{-6mm}
\end{figure}

In addition to the Butterfly model comparison, we explored the effectiveness of the sliding window attention versus FFT-based approximations in vision-specific tasks. Our analysis, detailed in Table~\ref{tab:vil-vs-pixelfly}, compares the accuracy of the state-of-the-art ViL (Vision Longformer) model~\cite{ViL}, which is supported by SWAT, against the SOTA FFT-attention-based Pixelfly model~\cite{pixelfly}. This comparison is particularly insightful as both models operate with a similar number of parameters. The results highlight that the ViL model achieves superior accuracy on the ImageNet-1K dataset, underscoring the effectiveness of sliding window attention in vision applications.

\vspace{-1.5mm}
\subsection{Performance comparison with Butterfly}
\vspace{-0.5mm}
Both SWAT and Butterfly accelerators were synthesized on FPGAs of the same characteristics\footnote{U55C (SWAT) and VCU128 (Butterfly) have the same number of logical resources}. They are also using a similar number of FPGA resources in FP16 precision, as shown in Table~\ref{tab:resource-usage}.
Using the cycle-accurate simulator provided by Butterfly, we independently evaluated the performance of the FFT-BTF engine and ATTN-BTF engine. As the original performance evaluation of Butterfly only considered the full-FFT configuration, we project its performance by computing the optimal ratio of resource distribution for FFT-BTF and ATTN-BTF engines at different input lengths.

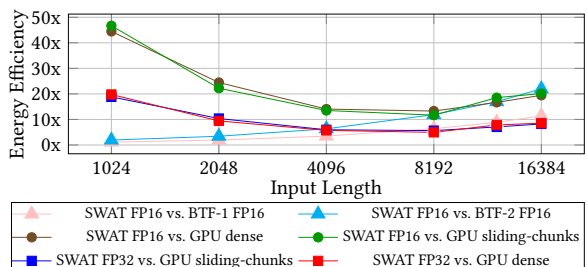
\begin{figure}[t]

\resizebox{.93\linewidth}{!}{
\begin{tikzpicture}

\begin{axis}[
  ylabel near ticks,
  %nodes near coords,
  ylabel={Energy Efficiency},
  width= 9.4cm,
  height=3.7cm,
  xmode=log,
  %legend style={font=\footnotesize, at={(1.3,0.9)},anchor=east,legend columns=2,
  legend style={font=\footnotesize, at={(1,-0.6)},anchor=east,legend columns=2,
            inner xsep=0pt,
            inner ysep=0pt, % Adjust vertical padding inside the legend
            outer xsep=0pt, % Adjust horizontal spacing outside the legend
            outer ysep=0pt, % Adjust vertical spacing outside the legend
},
  log ticks with fixed point,
  xtick={1024,2048,4096,8192,16384},
  xticklabels={1024,2048,4096,8192,16384},
x tick label style={rotate=0,anchor=north}, % This styles the tick labels
  ylabel shift=-5pt,
  xlabel={Input Length},
  xlabel style={yshift=2.5mm},
  ytick={50,40,30,20,10,0},
    title style={
    yshift=-2mm, % Moves the title 2mm closer
  },
  yticklabels={50x,40x,30x,20x,10x,0x},
  grid=major,
]

\addplot[color=pink, mark=triangle*, mark options={scale=1.7}] table[x=input_lenght, y=efficiency_BF1, col sep=space] {data/power-efficiency.data};
\addlegendentry{SWAT FP16 vs. BTF-1 FP16}

\addplot[color=cyan, mark=triangle*, mark options={scale=1.7}] table[x=input_lenght, y=efficiency_BF2, col sep=space] {data/power-efficiency.data};
\addlegendentry{SWAT FP16 vs. BTF-2 FP16}

\addplot[color=brown!60!black, mark=otimes*] table[x=input_lenght, y=efficiency_chunks_hf16, col sep=space] {data/power-efficiency.data};
\addlegendentry{SWAT FP16 vs. GPU dense}

\addplot[color=green!60!black, mark=otimes*] table[x=input_lenght, y=efficiency_dense_hf16, col sep=space] {data/power-efficiency.data};
\addlegendentry{SWAT FP16 vs. GPU sliding-chunks}

\addplot[color=blue, mark=square*] table[x=input_lenght, mark=diamond, y=efficiency_chunks, col sep=space] {data/power-efficiency.data};
\addlegendentry{SWAT FP32 vs. GPU sliding-chunks}

\addplot[color=red, mark=square*] table[x=input_lenght, y=efficiency_dense, col sep=space] {data/power-efficiency.data};
\addlegendentry{SWAT FP32 vs. GPU dense}

\end{axis}
    
\end{tikzpicture}
}
    \vspace{-1mm}
    \caption{Energy Efficiency of SWAT against SOTA GPU and FPGA implementations}
    \label{fig:power-efficiency}
\vspace{-6mm}
\end{figure}

SWAT's latency is based on its pipeline latency as stated earlier in Table~\ref{tab:pipeline-stages-timing}. 
For FPGA implementations, both accelerators produce consistent operation latencies regardless of the concrete values of input data, number of heads, layers, and batches. Total attention time is proportional to the execution time of a single head, which has an arbitrarily chosen dimensionality of 64. Figure~\ref{fig:performance} shows the speedup of SWAT in Longformer or Bigbird configuration against the Butterfly accelerator in BTF-1 (1 softmax layer) and BTF-2 (2 softmax layers) configurations across various sequence lengths, from 1024 to 16384 tokens.
Due to the poor scalability of the vanilla SoftMax attention, the Butterfly accelerator exhibits declining performance with long input sequences. At the standard Longformer configuration of 4096 input tokens (the accuracies in Table~\ref{tab:accuracy-LRA} are obtained at this configuration), SWAT performs 6.7$\times$ and 12.2$\times$ better respectively over BTF-1 and BTF-2. Due to the faster computation of SWAT, it also outperforms Butterfly from the energy perspective. We evaluate the power consumption of SWAT using the Xilinx Power Estimator. Figure~\ref{fig:power-efficiency} shows the energy efficiency (energy consumption per attention) of SWAT against BTF-1 and BTF-2. SWAT shows an increasing energy efficiency advantage along the input length, attaining 11.4$\times$ and 21.9$\times$ over BTF-1 and BTF-2 at 16384 context length, respectively.

\vspace{-1.5mm}
\subsection{Comparison with GPU}
\vspace{-0.5mm}
We benchmarked SWAT against GPU implementations with the same Transformer model, using AMD's rocBLAS and MIOpen libraries for tensor multiplication and SoftMax operation in the sliding chunks and the naïve dense approach which have been discussed in Section~\ref{sec:intro}.  We measured the execution time while excluding the overhead due to the first kernel launch and averaged the latency over 100 attention computations for consistency.
To compare fairly with GPU implementation, we synthesized an FP32 version of SWAT, which exhibits a higher pipeline latency of 264 cycles due to the FPGA's limitation on the FP32 MAC operation. 
The execution time comparison has been shown previously in Figure~\ref{fig:motivation}. At short input length, SWAT demonstrates better latency, which can be partly ascribed to the underutilization of the GPU in our single-batch experimental setup. However, as the input length reaches 4k, the GPU's execution time begins to rise sharply, indicating its full utilization. In contrast, SWAT exhibits a linearly increasing execution time in relation to input length, with similar performance of GPU between 4k and 8k input length but much better scalability for longer input length.  

A key aspect of SWAT is its energy efficiency, which is notably remarkable when compared to MI210, which has a power consumption of 300 watts. This efficiency is highlighted in Figure~\ref{fig:power-efficiency}, where SWAT's energy efficiency is compared against MI210 in both FP32 and FP16 precision. In FP32 precision, particularly considering the under-utilization of the GPU at shorter input lengths, SWAT achieves an impressive 20$\times$ energy efficiency advantage at an input length of 1k. However, as the GPU becomes better utilized with longer input sequences, SWAT's relative energy efficiency advantage decreases, reaching a minimum of 4.2$\times$ at an input length of 8k. SWAT's scalability, however, becomes increasingly pronounced with longer context lengths, and hence SWAT's superior energy efficiency grows, reaching up to approximately 8.4$\times$ that of GPU-based implementations at 16k input length.

\vspace{-1.5mm}
\section{Conclusion}
\vspace{-0.5mm}
We introduced SWAT, an FPGA-based accelerator specifically designed for window attention-based transformer models. Our approach is rooted in a comprehensive analysis of window attention workloads, leading to an input-stationary dataflow. This dataflow combines the advantages of FPGA's inherent distributed memory-and-computation architecture with a row-major dataflow and kernel fusion optimization. This unique combination effectively leverages the diagonal-structured sparsity inherent in sliding window attention, resulting in significantly improved performance. For long context lengths, SWAT stands out by delivering superior performance and energy efficiency over both the current SOTA FPGA-based static sparse attention accelerator and server-class GPUs.

\vspace{-1mm}
\bibliographystyle{ACM-Reference-Format}
\bibliography{refs}

\end{document}